\documentclass[conference]{IEEEtran}
\IEEEoverridecommandlockouts
\usepackage[top=2.3cm, bottom=3cm, left=1.92cm, right=1.92cm]{geometry}
\usepackage{cite}
\usepackage{amsmath,amssymb,amsfonts}
\usepackage{algorithm}
\usepackage{algorithmicx}
\usepackage{graphicx}
\usepackage{textcomp}
\usepackage{xcolor}
\usepackage{algpseudocode}
\usepackage{makecell}
\usepackage{url}
\usepackage{bbding}
\usepackage{subfig}
\usepackage{overpic}
\usepackage{float} 
\usepackage{tabularx}
\usepackage{xspace} 
\usepackage{graphicx}
\usepackage{bbding}
\usepackage{subfig}
\usepackage[numbers,sort&compress]{natbib}

\def\BibTeX{{\rm B\kern-.05em{\sc i\kern-.025em b}\kern-.08em
    T\kern-.1667em\lower.7ex\hbox{E}\kern-.125emX}}
\begin{document}

\title{Information Dissemination Model Based on User Attitude and Public Opinion Environment\\
}


\author{
	\IEEEauthorblockN{
		Xinyu Li\IEEEauthorrefmark{1}, 
		Jinyang Huang\IEEEauthorrefmark{1} \IEEEauthorrefmark{4},
		Xiang Zhang\IEEEauthorrefmark{2} \IEEEauthorrefmark{4},
        Peng Zhao\IEEEauthorrefmark{1},
        Meng Wang\IEEEauthorrefmark{1},
        \\Guohang Zhuang\IEEEauthorrefmark{1},
        Huan Yan\IEEEauthorrefmark{3},
		Xiao Sun\IEEEauthorrefmark{1}, 
		and Meng Wang\IEEEauthorrefmark{1}} 
	\IEEEauthorblockA{\IEEEauthorrefmark{1}School of Computer and Information, Hefei University of Technology, Hefei, China.}
	\IEEEauthorblockA{\IEEEauthorrefmark{2}CAS Key Laboratory of Electromagnetic Space Information, University of Science and Technology of China, Hefei, China.}
	\IEEEauthorblockA{\IEEEauthorrefmark{3} School of Big Data and Computer Science, Guizhou Normal University, Guiyang, China.} \IEEEauthorblockA{\IEEEauthorrefmark{4}Corresponding Authors: Jinyang Huang,  Xiang Zhang\quad Email: hjy@hfut.edu.cn, zhangxiang@ieee.org}
} 


\maketitle

\begin{abstract}
Modeling the information dissemination process in social networks is a challenging problem. Despite numerous attempts to address this issue, existing studies often assume that user attitudes have only one opportunity to alter during the information dissemination process. Additionally, these studies tend to consider the transformation of user attitudes as solely influenced by a single user, overlooking the dynamic and evolving nature of user attitudes and the impact of the public opinion environment. In this paper, we propose a novel model, \textit{UAPE}, which considers the influence of the aforementioned factors on the information dissemination process. Specifically, \textit{UAPE} regards the user's attitude towards the topic as dynamically changing, with the change jointly affected by multiple users simultaneously. Furthermore, the joint influence of multiple users can be considered as the impact of the public opinion environment. Extensive experimental results demonstrate that the model achieves an accuracy range of 91.62\% to 94.01\%, surpassing the performance of existing research.



\end{abstract}


\section{Introduction}

The growth of online social networking has minimized geographical constraints on information dissemination\cite{chen2021information}. However, the rapid spread of diverse, opinion-laden content remains a concern. This phenomenon can influence people's judgment to a certain extent and give rise to various social effects, such as affecting the fairness of elections, impacting the stock market, and even jeopardizing national stability. Therefore, accurately predicting the information dissemination trends in social networks plays a crucial role in scientifically responding to various emergent events \cite{9148985,9839133}.

Initially, Kempe et al. proposed \textit{the Independent Cascade (IC) Model} \cite{kempe2005influential} and \textit{the Linear Threshold (LT) Model} \cite{kempe2003maximizing} to predict the information dissemination trends. Later researchers found that apart from the neighboring nodes considered by the aforementioned models, additional factors also exert influence on the information dissemination process. Consequently, a substantial body of research has emerged, which incorporated different elements, e.g., user-interest topics \cite{qin2021influence,yu2020estimation, HjyPhD, zhang2020nsti}, individual preferences \cite{zhang2020nsti,wang2020topic,dai2022opinion}, information dissemination timing \cite{haldar2023temporal,9373931}, and user emotions \cite{hung2023cecm,wang2017emotion,huang2021node}, into \textit{IC} model.

However, previous research has ignored the dynamic nature of user attitudes and the influence of the public opinion environment, which greatly affects the prediction accuracy. Therefore, in this paper, we consider these two factors to simulate the dynamic process and trends of information dissemination to achieve a more accurate prediction. 

\begin{figure}[t]
\centering
\includegraphics[width=5.6cm, height = 4.5cm]{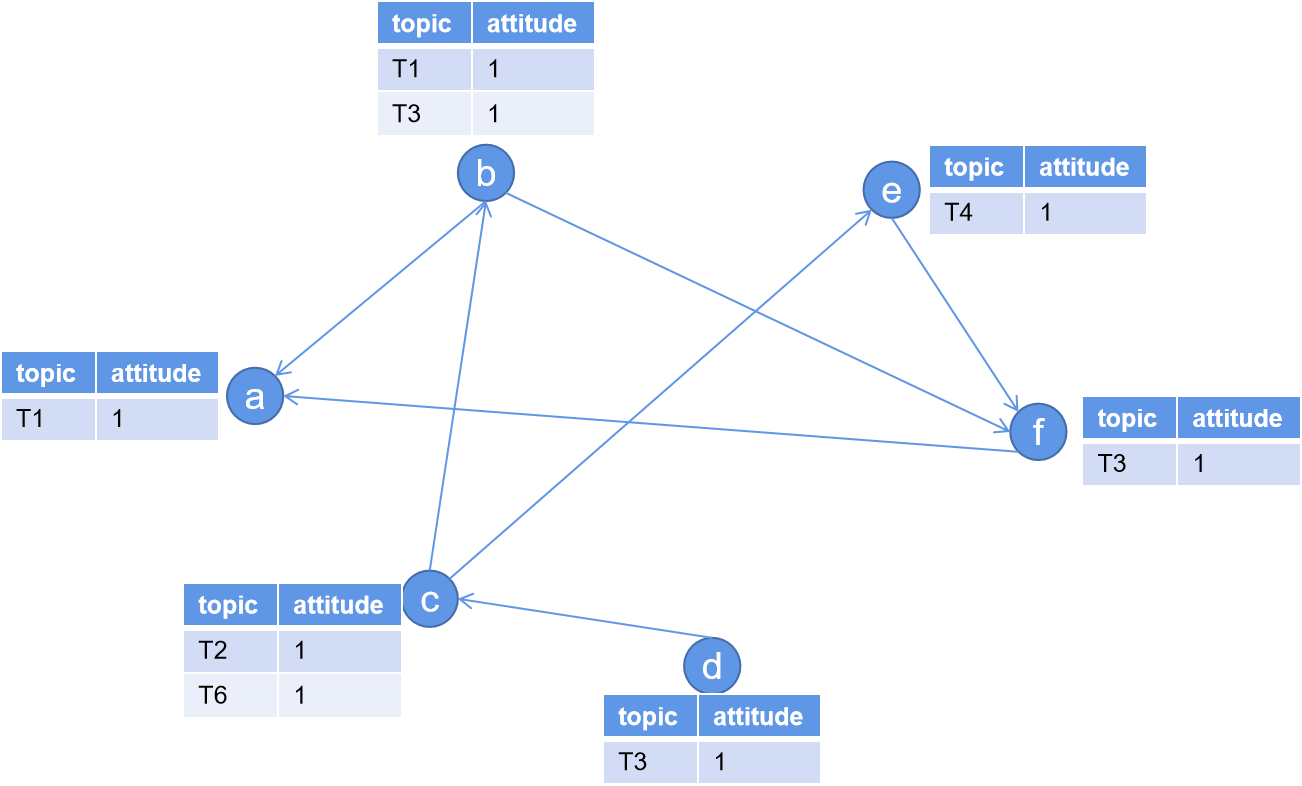}
\caption{Traditional Network Model.}
\end{figure}\label{fig:0}
\vspace{-0.1in}
\unskip





\section{Related Work}
A large number of studies have explored the information diffusion process. Kempe et al. proposed \textit{the Independent Cascade (IC) Model} \cite{kempe2005influential} and \textit{the Linear Threshold (LT) Model} \cite{kempe2003maximizing}. Considering the influence of topic, Qin et al. introduced a topic-aware community-based Model that incorporated both the structural characteristics of communities and thematic features \cite{qin2021influence}. \cite{yu2020estimation} considered the influence of similar topics between nodes and proposed a new model. \cite{zhang2020nsti} proposed \textit{NSTI-IC} model. \cite{tian2020deep} proposed a topic-aware model.

In addition to the user's interests and the topics they are interested in, individual factors, such as emotions, will also affect the information dissemination in social network. Therefore, considering the influence of individual factors, Zhang et al. introduced the \textit{NSTI-IC} model that incorporated the impact of individual preferences \cite{zhang2020nsti}. Wang et al. developed a unified probabilistic framework by considering users' historical emotional states, tweet topic distributions, and social structure \cite{wang2020topic}. Dai et al. proposed the \textit{EIC} model to elucidate the impact of group polarization effects and individual preferences \cite{dai2022opinion}. \cite{wang2017emotion} proposed an emotion-based independent cascade model that integrates user features, structural features, tweet features, and more. \cite{huang2021node} established an information dissemination model based on node sentiment.

Since the effective time of information dissemination can impact the dissemination range, taking this factor into account, \cite{haldar2023temporal} proposed the \textit{T-IC} model. Zhou et al. introduced a cycle-aware intelligent prediction method \cite{9373931}.



\section{Problem Formulation}

\begin{table*}[t]
\caption{FREQUENTLY USED NOTATION.}\label{tab1}
\centering
\begin{tabularx}{\textwidth}{ |c|
         >{\raggedright\arraybackslash}X|
        }
\hline
Notation & Description \\
\hline
$G=(\boldsymbol{V},\boldsymbol{E},\boldsymbol{T})$ & Represents a social network, where $\boldsymbol{V}$ is a set of nodes with a size of n, $\boldsymbol{E}$  is a set of edges with a size of m, and $\boldsymbol{T}$  is a set of topics with a size of z.\\
\hline
$P_e$ & Probability of information dissemination on edge e \cite{liu2021adaptive}.\\
\hline
$\boldsymbol{T_i}=<t_{i\_1},t_{i\_2},……,t_{i\_z}>$ & The set of user i attitudes towards all topics.\\
\hline
$sim(u,v)$ & Interest similarity between nodes u and v.\\
\hline
$A_{v}{i}$ & Attitude value of node v for topic i.\\
\hline
\end{tabularx}
\end{table*}

\begin{figure}[t]
\centering
\includegraphics[width=5.6 cm, height = 4.6cm]{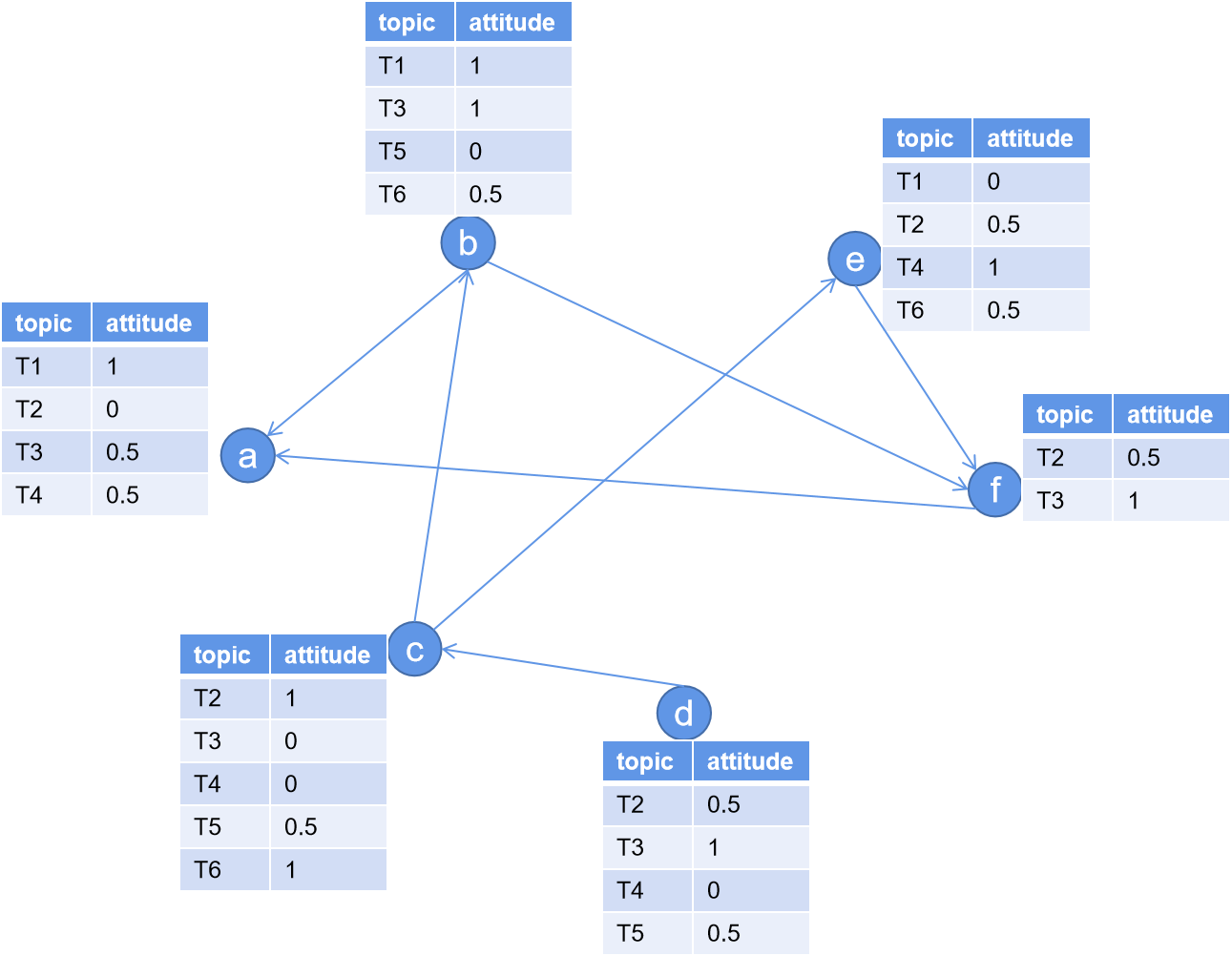}
\caption{Initial Network Model.}\label{fig:1}
\end{figure}   
\unskip


Unlike previous social network models, as shown in Fig.~\ref{fig:0}, which only consider the topics that users are initially interested in, ignoring the impact of other topics, we model the social network as a directed graph $G$ \cite{wang2023near}, as shown in Fig.~\ref{fig:1}, taking into account all topics with users' different attitudes. Each node $i$ is associated with a topic set $\boldsymbol{T_{i}}=\{t_{i\_1},t_{i\_2},t_{i\_3},……,t_{i\_k}\}$, reflecting the user's attitudes towards different topics. Specifically, $t_{i\_k} \in \{0,0.5, 1,-1\}$, and $t_{i\_k}$ is used to express the attitude of user $i$ towards topic $k$. The values $-1$, $0$, $0.5$, and $1$ indicate the user's stance, representing the four conditions: unknown, positive, neutral, and negative. Tab.~\ref{tab1} summarizes the commonly used symbols and their respective meanings\cite{10251628}.

\floatname{algorithm}{Algorithm}
\renewcommand{\algorithmicrequire}{\textbf{Input:}}
\renewcommand{\algorithmicensure}{\textbf{Output:}}

\begin{algorithm}
    \caption{UAPE}
        \begin{algorithmic}[1] 
            \Require $G=(\boldsymbol{V},\boldsymbol{E}, \boldsymbol{T})$, initial set of topics $\boldsymbol{T_{k}}=\{\boldsymbol{T_{1}},\boldsymbol{T_{2}},……,\boldsymbol{T_{z}}\}$,initial topic number j,Time threshold K.
            \Ensure  The updated network G=(V,E,T).
            \Function {UAPE}{$G,\boldsymbol{T_{k}}, j, K$}
            \State $\boldsymbol{V_{j}} = \boldsymbol{T_{j-0}} \cup \boldsymbol{T_{j-0.5}} \cup \boldsymbol{T_{j-1}}$
            \State $\boldsymbol{V_{adj}} = \boldsymbol{\emptyset}$, $\boldsymbol{S} = \boldsymbol{\emptyset}$
                \For{k = 1 to K}
                    \For{v in $\boldsymbol{V_{j}}$}
                        \State $\boldsymbol{S}=\{u|(u,v)\in \boldsymbol{E}\}$
                        \For{q $\in$ $\boldsymbol{S}$}
                            \State $T_{cur} = T_{q}^{j}$
                            \State $A_{q}^{j} = Eq.(2)$
                            \State $T_{q}^{j} = Get\_Att(G,q,v,j,A_{q}^{j})$
                            \If{q $\notin \boldsymbol{V_{adj}}$}
                                \State $\boldsymbol{V_{adj}} \gets q$
                            \EndIf
                            \If{$T_{cur} = -1$ and $T_{q}^{j} != -1$}
                                \State $\boldsymbol{T_{j-T_{q}^{j}}} \gets q$
                                \State $\boldsymbol{V_{j}} \gets q$
                            \EndIf
                        \EndFor
                    \EndFor
                \EndFor
                \State \Return{$G$}
            \EndFunction
        \end{algorithmic}
\end{algorithm}

\floatname{algorithm}{Algorithm}
\renewcommand{\algorithmicrequire}{\textbf{Input:}}
\renewcommand{\algorithmicensure}{\textbf{Output:}}

\begin{algorithm}
    \caption{$Get\_Att$}
        \begin{algorithmic}[1] 
            \Require $G=(\boldsymbol{V},\boldsymbol{E}, \boldsymbol{T})$, node q,v, the topic number j,$A_{q}^{j}$.
            \Ensure  $T_{q}^{j}$.
            \Function {$Get\_Att$}{$G,q,v,j,A_{q}^{j}$}
            \State $T_{cur} = T_{q}^{j}$
             \If{$T_{q}^{j}=0.5$ or -1}
                \If{$P_{j}(q,v) > A_{q}^{j}$}
                    \State $T_{q}^{j} = T_{v}^{j}$
                \Else
                    \State $T_{q}^{j} = 0.5$
                \EndIf
            \Else
                \State $T_{q}^{j} = (\overline{T_{q}^{j} \oplus T_{v}^{j}}) \ast  T_{q}^{j} + (T_{q}^{j} \oplus T_{v}^{j}) \ast  (T_{q}^{j} \pm \varepsilon \ast 0.5)$
            \EndIf
            \If{$T_{cur} != T_{q}^{j}$}
                \State $\boldsymbol{T_{j-T_{q}^{j}}} \gets q$
                \State $\boldsymbol{T_{j-T_{cur}}} = \boldsymbol{T_{j-T_{cur}}} - q$
            \EndIf
            \State \Return{$T_{q}^{j}$}
            \EndFunction
        \end{algorithmic}
\end{algorithm}

\begin{figure*}[h]
	\centering
	\subfloat[Dataset \uppercase\expandafter{\romannumeral1} Experimental results.]{\includegraphics[width=0.33\textwidth]{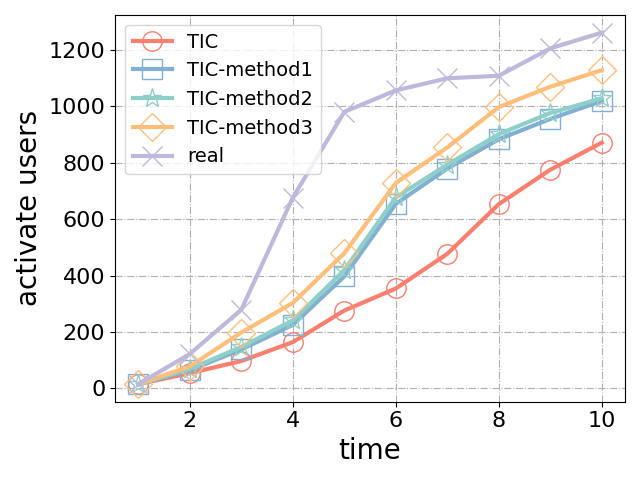}}
	\subfloat[Dataset \uppercase\expandafter{\romannumeral2} Experimental results.]{\includegraphics[width=0.33\textwidth]{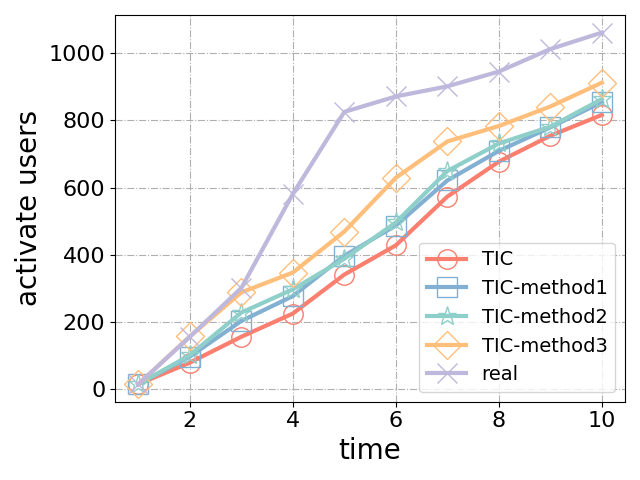}}
	\subfloat[Dataset \uppercase\expandafter{\romannumeral3} Experimental results.]{\includegraphics[width=0.33\textwidth]{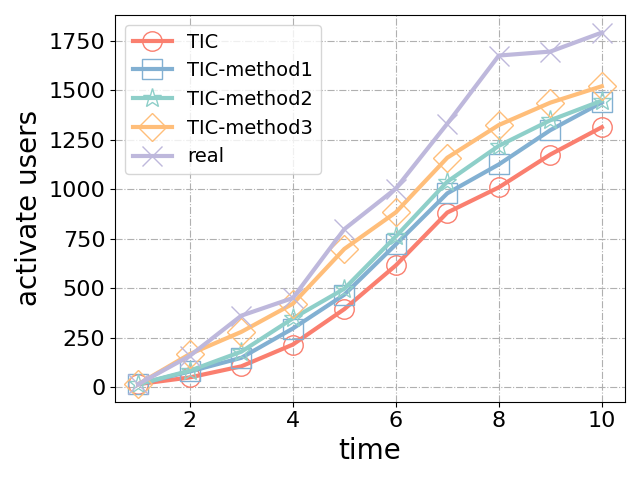}}\\
    \subfloat[Dataset \uppercase\expandafter{\romannumeral4} Experimental results.]{\includegraphics[width=0.33\textwidth]{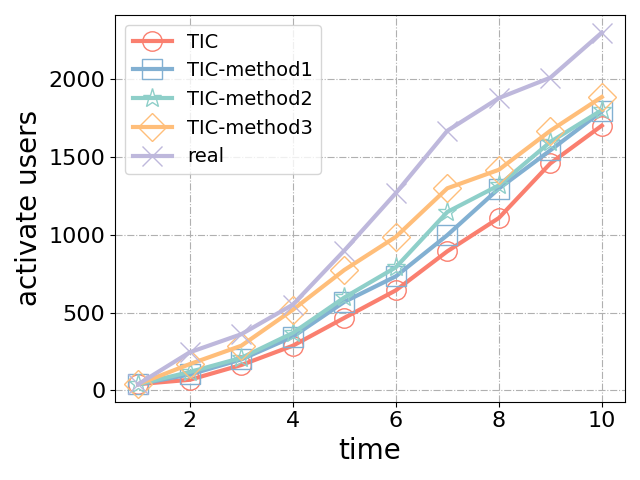}}
	\subfloat[Dataset \uppercase\expandafter{\romannumeral5} Experimental results.]{\includegraphics[width=0.33\textwidth]{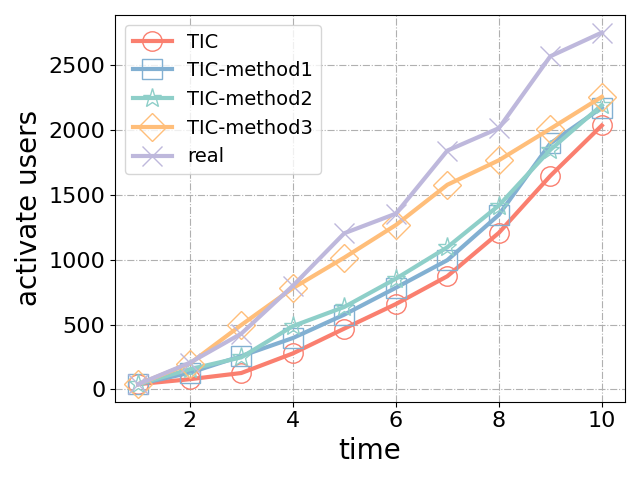}}
	\subfloat[Dataset \uppercase\expandafter{\romannumeral6} Experimental results.]{\includegraphics[width=0.33\textwidth]{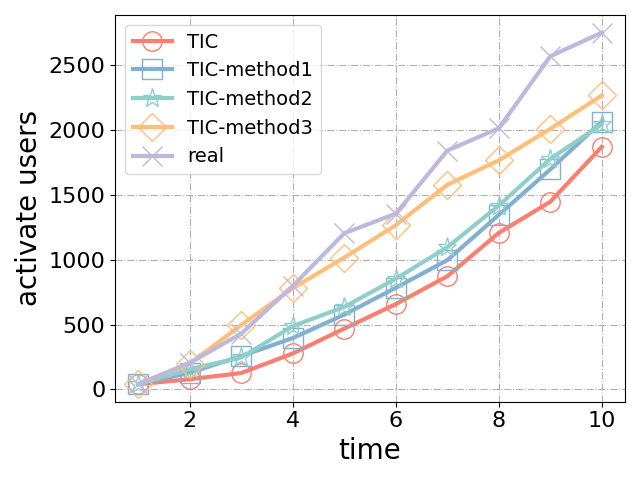}}
	\caption{Experimental results of \textit{TIC} model.}
    \label{fig2}
\end{figure*}

\begin{figure*}[h]
	\centering
	\subfloat[Dataset \uppercase\expandafter{\romannumeral1} Experimental results.]{\includegraphics[width=0.33\textwidth]{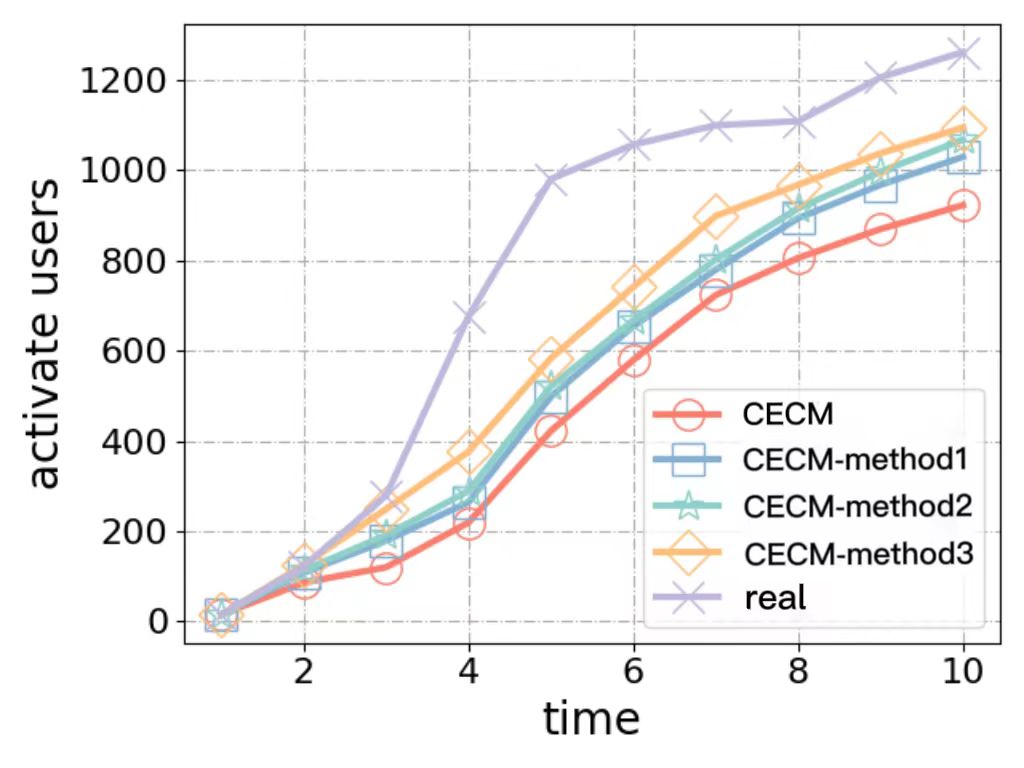}}
	\subfloat[Dataset \uppercase\expandafter{\romannumeral2} Experimental results.]{\includegraphics[width=0.33\textwidth]{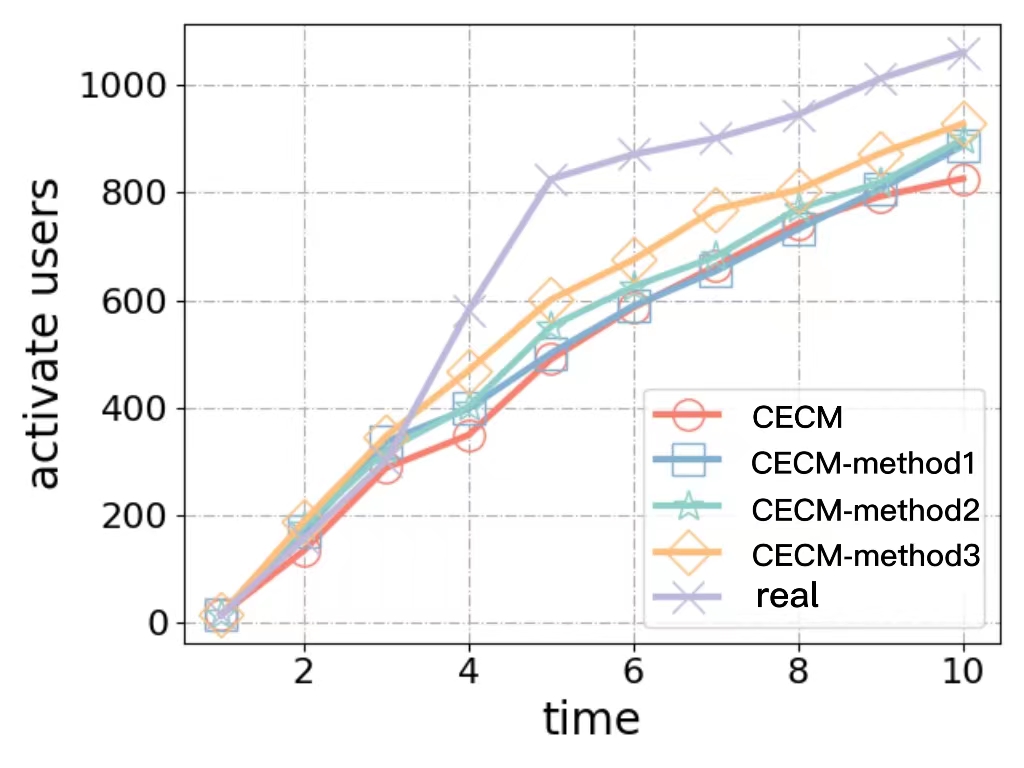}}
	\subfloat[Dataset \uppercase\expandafter{\romannumeral3} Experimental results.]{\includegraphics[width=0.33\textwidth]{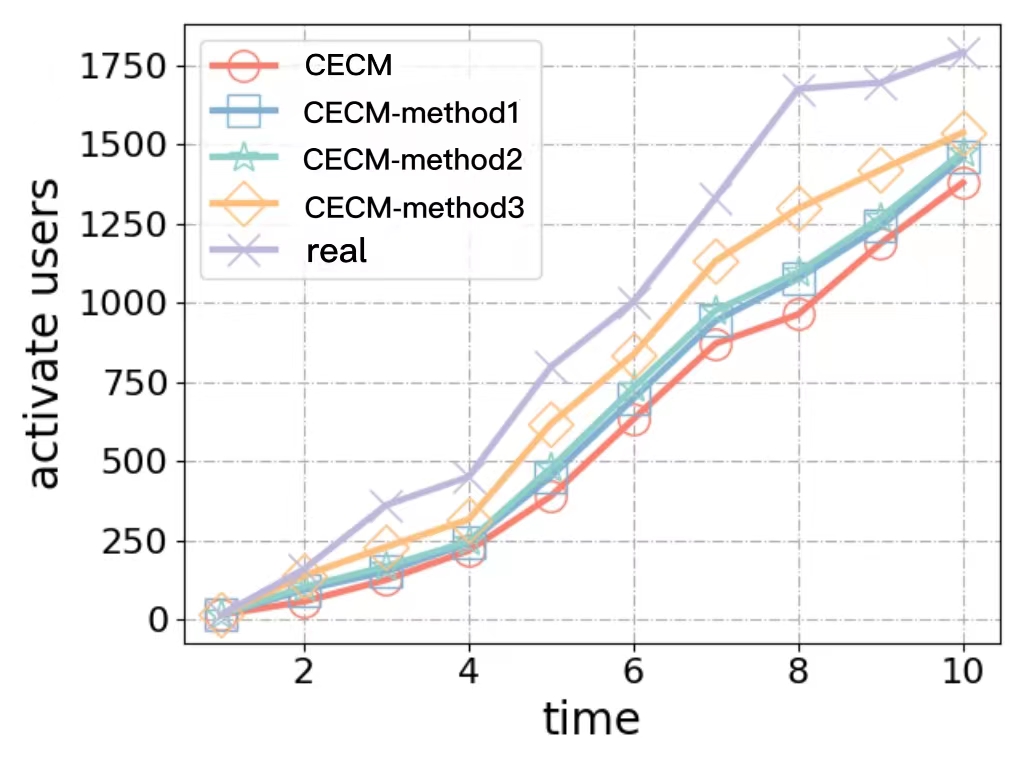}}\\
    \subfloat[Dataset \uppercase\expandafter{\romannumeral4} Experimental results.]{\includegraphics[width=0.33\textwidth]{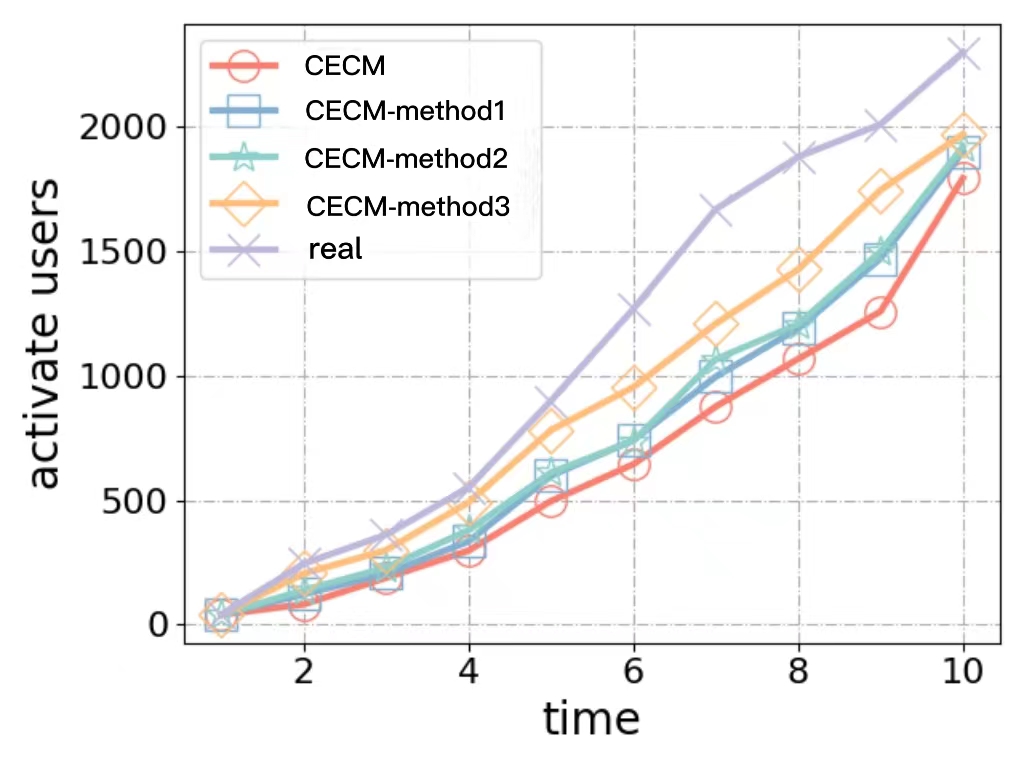}}
	\subfloat[Dataset \uppercase\expandafter{\romannumeral5} Experimental results.]{\includegraphics[width=0.33\textwidth]{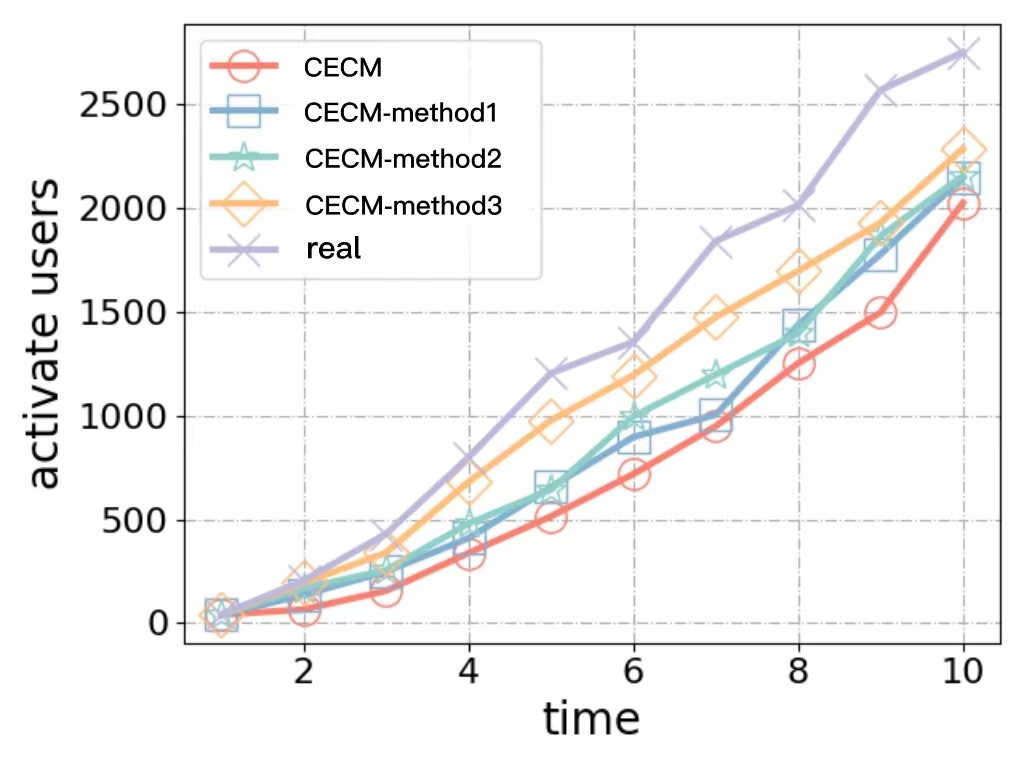}}
	\subfloat[Dataset \uppercase\expandafter{\romannumeral6} Experimental results.]{\includegraphics[width=0.33\textwidth]{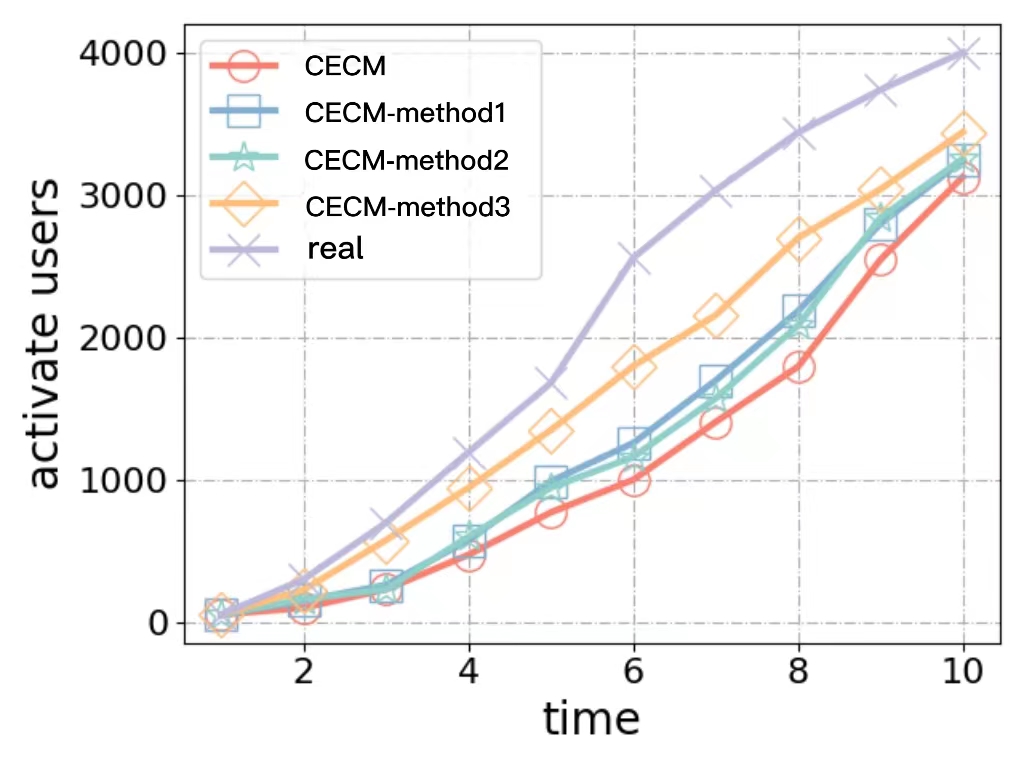}}
	\caption{Experimental results of \textit{CECM} model.}
    \label{fig3}
\end{figure*}

\begin{figure*}[h]
	\centering
	\subfloat[Dataset \uppercase\expandafter{\romannumeral1} Experimental results.]{\includegraphics[width=0.33\textwidth]{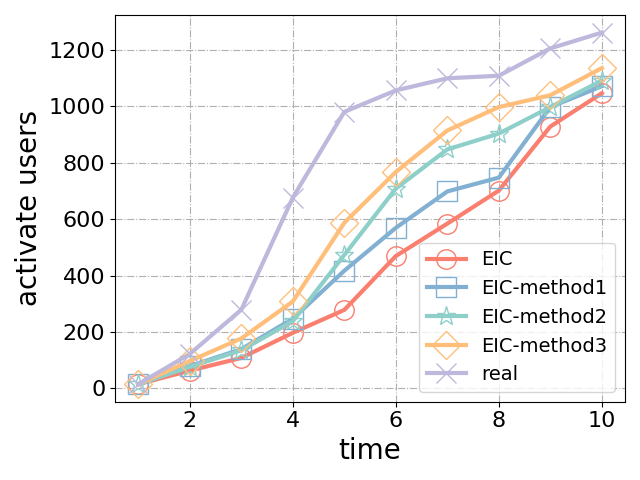}}
	\subfloat[Dataset \uppercase\expandafter{\romannumeral2} Experimental results.]{\includegraphics[width=0.33\textwidth]{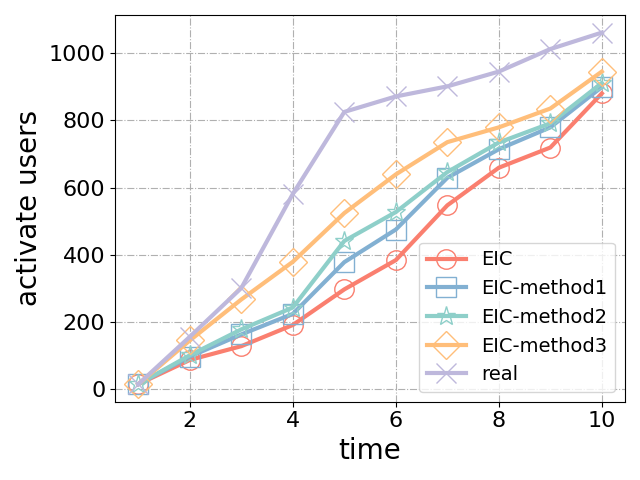}}
	\subfloat[Dataset \uppercase\expandafter{\romannumeral3} Experimental results.]{\includegraphics[width=0.33\textwidth]{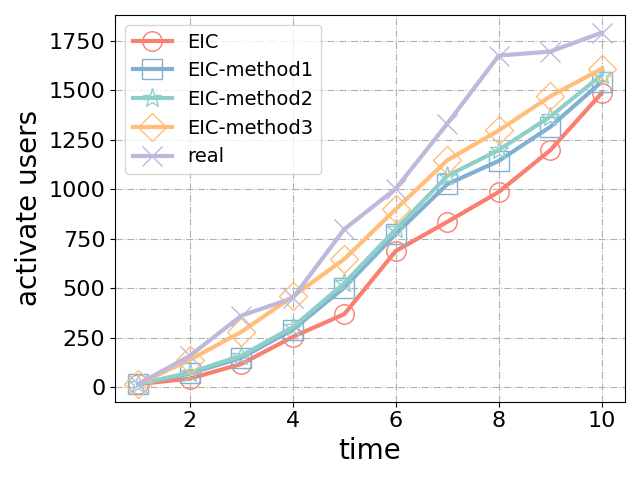}}\\
    \subfloat[Dataset \uppercase\expandafter{\romannumeral4} Experimental results.]{\includegraphics[width=0.33\textwidth]{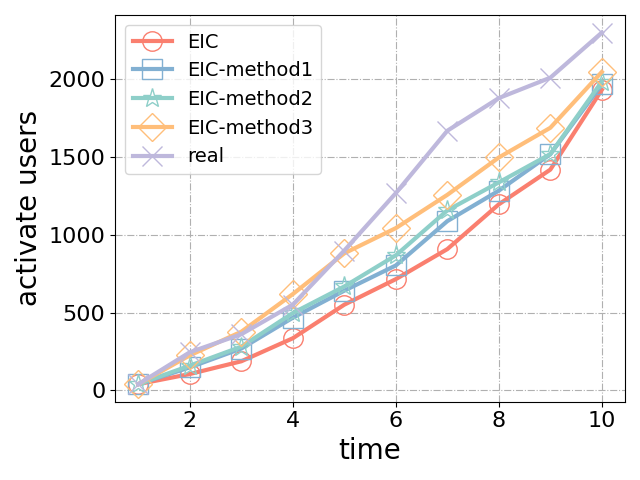}}
	\subfloat[Dataset \uppercase\expandafter{\romannumeral5} Experimental results.]{\includegraphics[width=0.33\textwidth]{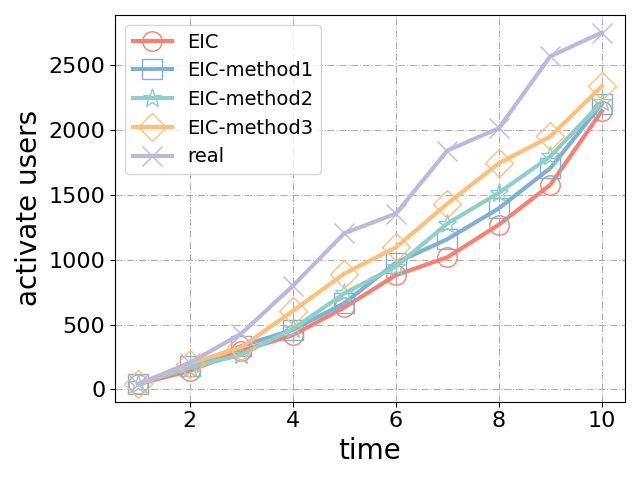}}
	\subfloat[Dataset \uppercase\expandafter{\romannumeral6} Experimental results.]{\includegraphics[width=0.33\textwidth]{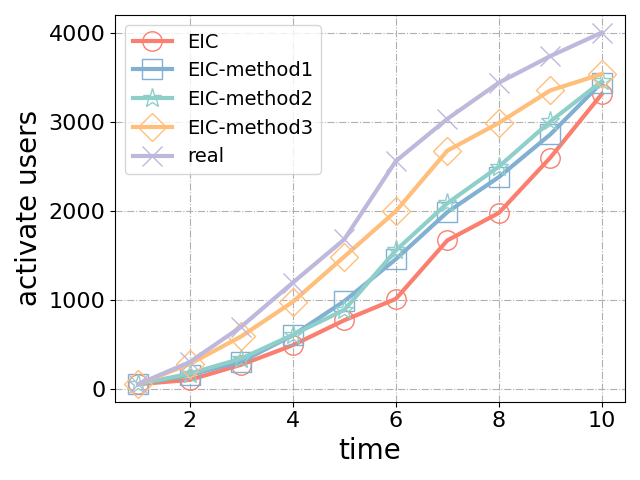}}
	\caption{Experimental results of \textit{EIC} model.}
    \label{fig4}
\end{figure*}

\subsection{Influence Evaluation Mechanism}
Given two nodes $u$ and $v$, the topic sets corresponding to these two nodes are $\boldsymbol{T_{u}}=\{t_{u\_1},t_{u\_2},t_{u\_3},……, t_{u\_n}\}$ and $\boldsymbol{T_{v}}=\{t_{v\_1},t_{v\_2},t_{v\_3},……,t_{v\_m}\}$. We employ the Degrootian model \cite{degroot1974reaching} to calculate the changes in users' attitudes toward each topic during information dissemination:

\begin{equation}
O_{i\_v}(t+1) = \frac{T_{i\_v}(t) + \sum_{j \in T_{i}(t)} T_{j\_v}(t)}{|T_{i}(t)| + 1}
\label{eq:3}
\end{equation}

For topic $i$, the influence $P_i$ of node $u$ on node $v$ is measured as follows:
{\setlength\abovedisplayskip{0pt}
\setlength\belowdisplayskip{3pt}
\begin{equation}
Pr\left\{v \rightarrow i\right\} = \frac{\frac{1}{|O_{i\_v}(t) - O_{i}| + 0.01}}{\sum_{j \in T_{i}(t) \frac{1}{|O_{i\_v}(t) - O_{j}| + 0.01}}}
\label{eq:1}
    \end{equation}

\begin{equation}
\begin{aligned}
P_{i}(u,v) = Pr\left\{v \rightarrow i\right\} \ast \sqrt{\sum_{i=1}^k\Big(\big( t_{u}^{i} - t_{v}^{i}\big) \oplus 0.5\Big)^2}/IN(v)
\label{eq:2}
\end{aligned}
\end{equation}
 where $Pr\left\{v \rightarrow i\right\}$ is used to calculate the level of interest of user $v$ in topic $i$. The denominator 0.01 is utilized to handle exceptional cases, a similar approach to the one employed in the study \cite{kozitsin2020users}. The variable $k$ represents the size of the set of topics in which both nodes $u$ and $v$ are interested, and $IN(v)$ represents the in-degree of node $v$.

\subsection{The Attitude Persistence of A Node to A Topic}
For every topic $t_i$ within the social network, each node holds a specific attitude towards it. During each iteration, the information received by a node regarding the same topic may contain varying attitudes. These diverse attitudes can influence the node's persistence in its attitude, potentially leading to changes in the node's attitude. Therefore, in this paper, we propose a computational system that quantifies the persistence of nodes to their respective attitudes, providing insights into the likelihood of attitude changes concerning topic $t_i$.

\begin{equation}
A_{v}^{i} = A_{v}^{i} - \sum_{u=1}^q\Big(|t_{u}^{i} - t_{v}^{i}| \ast P_{i}(u,v) - (\overline{t_{u}^{i} \oplus t_{v}^{i}}) \ast P_{i}(u,v) \Big)/q
\label{eq:2}
\end{equation}
where, $q$ represents the number of messages received by node $v$ so far, and $A_{v}^{i}$ represents node $v$'s persistence of attitude towards topic $t_i$. 

\section{Algorithm}
\subsection{Information dissemination Model}
Based on the \textit{IC} model and considering the dynamic changes in user attitudes during the information dissemination process, along with the combined influence of multiple neighboring users, we present the \textit{UAPE} model. The detailed process of the model is outlined in Algorithm $1$. Where $V_{adj}$ denotes the set of adjacent nodes of all nodes in the existing seed set $S$, $T_{cur}$ represents the state of the node before it is affected.


\subsection{Node Attitude Change Process}
During the process of information dissemination\cite{10339891} in social networks, the attitudes of nodes towards topics undergo continuous changes. Therefore, it is crucial to track the evolving attitudes of nodes towards various topics throughout the information dissemination process. To achieve this, we propose an algorithm $Get\_Att$, that comprehensively considers all the information received by the node about the specified topic up to the current moment and calculates the node's attitude towards the specified topic. The detailed process of the model is outlined in Algorithm $2$.



\section{Experiment}
\subsection{Dataset}

The datasets used in this study are all obtained from real Weibo data. Datasets \uppercase\expandafter{\romannumeral1},\uppercase\expandafter{\romannumeral2} and \uppercase\expandafter{\romannumeral3} contain original tweets and comments related to a particular topic, respectively. Datasets \uppercase\expandafter{\romannumeral4} and \uppercase\expandafter{\romannumeral5} contain original tweets on two specific topics and comments related to them. Dataset \uppercase\expandafter{\romannumeral6} contains all original Weibo posts and comments on three topics. Tab.~\ref{tab2} provides detailed information about the dataset.


\subsection{Comparison Methods}
The \textit{UAPE} model is compared with the \textit{TIC} model, which solely considered topics in reference \cite{barbieri2013topic}. Additionally, we compare it with the \textit{EIC} \cite{dai2022opinion} and the \textit{CECM} \cite{hung2023cecm} model, both of which considered user sentiment or individual preferences. Specifically, the methods evaluated and compared in the experiments are as follows:


\begin{itemize}
\item 
Method 1: Considering the dynamic changes in user attitudes during the information dissemination process to make predictions.
\item 
Method 2: Taking into account the collective influence of multiple neighboring users to make predictions.
\item 
Method 3: Considering the dynamic changes in user attitudes during the information dissemination process and the combined influence of multiple neighboring users to make predictions.
\end{itemize}

\begin{table}[t]
\caption{The Details of The Dataset.}\label{tab2}
\begin{center}
\begin{tabular}{|c |c |c |c |c|}
\hline
\textbf{Dataset} & \textbf{nodes} & \textbf{edges} & \textbf{topics} & \textbf{Number of seed nodes} \\
\hline
\textbf{Dataset \uppercase\expandafter{\romannumeral1}}  & 1331 & 8737 & 1 & 20\\
\hline
\textbf{Dataset \uppercase\expandafter{\romannumeral2}} & 1109 & 7723 & 1 & 23\\
\hline
\textbf{Dataset \uppercase\expandafter{\romannumeral3}} & 1801 & 8493 & 1 & 27\\
\hline
\textbf{Dataset \uppercase\expandafter{\romannumeral4}} & 2351 & 14739 & 2 & 41\\
\hline
\textbf{Dataset \uppercase\expandafter{\romannumeral5}} & 2817 & 13283 & 2 & 45\\
\hline
\textbf{Dataset \uppercase\expandafter{\romannumeral6}} & 4028 & 23151 & 3 & 69\\
\hline
\end{tabular}
\end{center}
\end{table}

\begin{table}[t]
\caption{AUC values for \textit{TIC}, \textit{CECM}, \textit{EIC} and \textit{UAPE}. }\label{tab3}
\begin{center}
\begin{tabular}{|c| c |c |c |c |}
\hline
\textbf{Dataset} & \textbf{TIC} & \textbf{CECM} & \textbf{EIC} &  \textbf{UAPE} \\
\hline
\textbf{Dataset \uppercase\expandafter{\romannumeral1}} & 77.01$\%$ & 80.63$\%$ & 84.48$\%$ & 93.90$\%$ \\
\hline
\textbf{Dataset \uppercase\expandafter{\romannumeral2}} & 76.92$\%$ & 79.92$\%$ & 84.64$\%$ & 93.71$\%$ \\
\hline
\textbf{Dataset \uppercase\expandafter{\romannumeral3}} & 76.68$\%$ & 80.01$\%$ & 84.97$\%$ & 92.82$\%$ \\
\hline
\textbf{Dataset \uppercase\expandafter{\romannumeral4}} & 75.49$\%$ & 80.15$\%$ & 83.12$\%$ & 94.01$\%$\\
\hline
\textbf{Dataset \uppercase\expandafter{\romannumeral5}} & 75.97$\%$ & 79.69$\%$ & 82.89$\%$ & 93.67$\%$\\
\hline
\textbf{Dataset \uppercase\expandafter{\romannumeral6}} & 74.82$\%$ & 79.43$\%$ & 82.77$\%$ & 91.62$\%$ \\
\hline
\end{tabular}
\end{center}
\end{table}

\subsection{Experimental Results}

$(a)$ to $(f)$ of Fig.~\ref{fig2}, Fig.~\ref{fig3}, and Fig.~\ref{fig4} represent the \textit{TIC}, the \textit{CECM}, and the \textit{EIC} model on the six datasets, respectively. We compare the prediction process of our model with the actual information dissemination process. It is evident that both the dynamic changes in user attitudes during information dissemination and the joint effect of multiple neighboring users significantly affect information dissemination.


In addition, we conduct experiments on the \textit{UAPE} model proposed in this paper and compare its performance with that of the \textit{TIC}, \textit{CECM}, and \textit{EIC} models. The experimental results, as depicted in Tab.~\ref{tab3}, clearly demonstrate a substantial improvement in prediction accuracy achieved by our model in comparison to the other models.

\section{CONCLUSION}

In this paper, we propose a novel model, \textit{UAPE}, which considers the impacts of dynamically changing user attitudes and the public opinion environment on the information dissemination process to accurately depict information dissemination in online social networks. Specifically, dynamically changing user attitudes and the joint effects of multiple neighboring users are considered essential reference indicators to realize the accurate prediction of the information dissemination trend. Extensive experiments demonstrate that \textit{UAPE} can achieve more satisfactory prediction accuracy on the dissemination trend.


\section{Acknowledge}
This work is supported by the National Natural Science Foundation of China (Grant No. 62302145), the key program of Anhui Province Key Laboratory of Affective Computing and Advanced Intelligent Machine (Grant No. PA2023GDSK0059), and Young teachers’ scientific research innovation launches special A project (Grant No. JZ2023HGQA0100). 

\bibliographystyle{IEEEtran}
\bibliography{con}

\begin{thebibliography}{10}
\providecommand{\url}[1]{#1}
\csname url@samestyle\endcsname
\providecommand{\newblock}{\relax}
\providecommand{\bibinfo}[2]{#2}
\providecommand{\BIBentrySTDinterwordspacing}{\spaceskip=0pt\relax}
\providecommand{\BIBentryALTinterwordstretchfactor}{4}
\providecommand{\BIBentryALTinterwordspacing}{\spaceskip=\fontdimen2\font plus
\BIBentryALTinterwordstretchfactor\fontdimen3\font minus \fontdimen4\font\relax}
\providecommand{\BIBforeignlanguage}[2]{{%
\expandafter\ifx\csname l@#1\endcsname\relax
\typeout{** WARNING: IEEEtran.bst: No hyphenation pattern has been}%
\typeout{** loaded for the language `#1'. Using the pattern for}%
\typeout{** the default language instead.}%
\else
\language=\csname l@#1\endcsname
\fi
#2}}
\providecommand{\BIBdecl}{\relax}
\BIBdecl

\bibitem{chen2021information}
X.~Chen, C.~Wu, T.~Chen, Z.~Liu, H.~Zhang, M.~Bennis, H.~Liu, and Y.~Ji, ``Information freshness-aware task offloading in air-ground integrated edge computing systems,'' \emph{IEEE Journal on Selected Areas in Communications}, vol.~40, no.~1, pp. 243--258, 2021.

\bibitem{9148985}
R.~A. Banez, H.~Gao, L.~Li, C.~Yang, Z.~Han, and H.~V. Poor, ``Belief and opinion evolution in social networks based on a multi-population mean field game approach,'' in \emph{ICC 2020 - 2020 IEEE International Conference on Communications (ICC)}, 2020, pp. 1--6.

\bibitem{9839133}
N.~Niknami and J.~Wu, ``A budged framework to model a multi-round competitive influence maximization problem,'' in \emph{ICC 2022 - IEEE International Conference on Communications}, 2022, pp. 4120--4125.

\bibitem{kempe2005influential}
D.~Kempe, J.~M. Kleinberg, and {\'E}.~Tardos, ``Influential nodes in a diffusion model for social networks.'' in \emph{ICALP}, vol.~5.\hskip 1em plus 0.5em minus 0.4em\relax Springer, 2005, pp. 1127--1138.

\bibitem{kempe2003maximizing}
D.~Kempe, J.~Kleinberg, and {\'E}.~Tardos, ``Maximizing the spread of influence through a social network,'' in \emph{Proceedings of the ninth ACM SIGKDD international conference on Knowledge discovery and data mining}, 2003, pp. 137--146.

\bibitem{qin2021influence}
X.~Qin, C.~Zhong, and Q.~Yang, ``An influence maximization algorithm based on community-topic features for dynamic social networks,'' \emph{IEEE Transactions on Network Science and Engineering}, vol.~9, no.~2, pp. 608--621, 2021.

\bibitem{yu2020estimation}
M.~Yu, V.~Gupta, and M.~Kolar, ``Estimation of a low-rank topic-based model for information cascades,'' \emph{The Journal of Machine Learning Research}, vol.~21, no.~1, pp. 2721--2767, 2020.

\bibitem{HjyPhD}
J.~Huang, ``Research on anti-interference wifi-based human activity recognition method,'' Ph. D. dissertation, University of Science and Technology of China, 2022, doi:10.27517/d.cnki.gzkju.2022.000757.

\bibitem{zhang2020nsti}
C.~Zhang, Y.~Yin, and Y.~Liu, ``Nsti-ic: An independent cascade model based on neighbor structures and topic-aware interest,'' in \emph{Web and Big Data: 4th International Joint Conference, APWeb-WAIM 2020, Tianjin, China, September 18-20, 2020, Proceedings, Part I 4}.\hskip 1em plus 0.5em minus 0.4em\relax Springer, 2020, pp. 170--178.

\bibitem{wang2020topic}
X.~Wang, D.~Jin, K.~Musial, and J.~Dang, ``Topic enhanced sentiment spreading model in social networks considering user interest,'' in \emph{Proceedings of the AAAI Conference on Artificial Intelligence}, vol.~34, no.~01, 2020, pp. 989--996.

\bibitem{dai2022opinion}
J.~Dai, J.~Zhu, and G.~Wang, ``Opinion influence maximization problem in online social networks based on group polarization effect,'' \emph{Information Sciences}, vol. 609, pp. 195--214, 2022.

\bibitem{haldar2023temporal}
A.~Haldar, S.~Wang, G.~V. Demirci, J.~Oakley, and H.~Ferhatosmanoglu, ``Temporal cascade model for analyzing spread in evolving networks,'' \emph{ACM Transactions on Spatial Algorithms and Systems}, 2023.

\bibitem{9373931}
X.~Zhou, W.~Liang, Z.~Luo, and Y.~Pan, ``Periodic-aware intelligent prediction model for information diffusion in social networks,'' \emph{IEEE Transactions on Network Science and Engineering}, vol.~8, no.~2, pp. 894--904, 2021.

\bibitem{hung2023cecm}
C.-C. Hung, X.~Gao, Z.~Liu, Y.~Chai, T.~Liu, and C.~Liu, ``Cecm: A cognitive emotional contagion model in social networks,'' \emph{Multimedia Tools and Applications}, pp. 1--23, 2023.

\bibitem{wang2017emotion}
Q.~Wang, Y.~Jin, T.~Yang, and S.~Cheng, ``An emotion-based independent cascade model for sentiment spreading,'' \emph{Knowledge-Based Systems}, vol. 116, pp. 86--93, 2017.

\bibitem{huang2021node}
H.~Huang, T.~Wang, M.~Hu, M.~Dong, and L.~Lai, ``Node attitude aware information dissemination model based on evolutionary game in social networks,'' \emph{Mobile Networks and Applications}, vol.~26, pp. 114--129, 2021.

\bibitem{tian2020deep}
S.~Tian, S.~Mo, L.~Wang, and Z.~Peng, ``Deep reinforcement learning-based approach to tackle topic-aware influence maximization,'' \emph{Data Science and Engineering}, vol.~5, pp. 1--11, 2020.

\bibitem{liu2021adaptive}
J.~Liu, H.~Xu, L.~Wang, Y.~Xu, C.~Qian, J.~Huang, and H.~Huang, ``Adaptive asynchronous federated learning in resource-constrained edge computing,'' \emph{IEEE Transactions on Mobile Computing}, 2021.

\bibitem{wang2023near}
X.~Wang, Z.~Liu, A.~X. Liu, X.~Zheng, H.~Zhou, A.~Hawbani, and Z.~Dang, ``A near-optimal protocol for continuous tag recognition in mobile rfid systems,'' \emph{IEEE/ACM Transactions on Networking}, 2023.

\bibitem{10251628}
J.~Liu, J.~Yan, H.~Xu, Z.~Wang, J.~Huang, and Y.~Xu, ``Finch: Enhancing federated learning with hierarchical neural architecture search,'' \emph{IEEE Transactions on Mobile Computing}, pp. 1--15, 2023.

\bibitem{degroot1974reaching}
M.~H. DeGroot, ``Reaching a consensus,'' \emph{Journal of the American Statistical association}, vol.~69, no. 345, pp. 118--121, 1974.

\bibitem{kozitsin2020users}
I.~V. Kozitsin and A.~G. Chkhartishvili, ``Users’ activity in online social networks and the formation of echo chambers,'' in \emph{2020 13th International Conference" Management of large-scale system development"(MLSD)}.\hskip 1em plus 0.5em minus 0.4em\relax IEEE, 2020, pp. 1--5.

\bibitem{10339891}
J.~Huang, B.~Liu, C.~Miao, X.~Zhang, J.~Liu, L.~Su, Z.~Liu, and Y.~Gu, ``Phyfinatt: An undetectable attack framework against phy layer fingerprint-based wifi authentication,'' \emph{IEEE Transactions on Mobile Computing}, pp. 1--18, 2023.

\bibitem{barbieri2013topic}
N.~Barbieri, F.~Bonchi, and G.~Manco, ``Topic-aware social influence propagation models,'' \emph{Knowledge and information systems}, vol.~37, pp. 555--584, 2013.

\end{thebibliography}

\end{document}